

\tolerance=10000
\input phyzzx.tex

\def\W {{\cal W }}
\nopubblock

{\begingroup \tabskip=\hsize minus \hsize
   \baselineskip=1.5\ht\strutbox \topspace-2\baselineskip
   \halign to\hsize{\strut #\hfil\tabskip=0pt\crcr
   {QMW-92-21}\cr {NI-92021}\cr {hep-th/9301074} \cr
   {December 1992}\cr }\endgroup}

\titlepage
\title {{\bf  GEOMETRY AND \W -GRAVITY}\foot{Talk given
at {\it Pathways to Fundamental Interactions}, the
16th John Hopkins Workshop on Current Problems in Particle Theory, Gothenborg,
1992.}}

\author {C. M. Hull}
\
\address {Physics Department,
Queen Mary and Westfield College,
\break
Mile End Road, London E1 4NS, United Kingdom.}

 \andaddress{Isaac Newton Institute, 20 Clarkson Road,
 \break
 Cambridge  CB2 0EH,
 United Kingdom.}

\abstract
{The higher-spin geometries of   $\W_\infty$-gravity and
 $\W_N$-gravity are analysed and used to derive the
complete non-linear structure of the coupling to matter and
its symmetries.  The
symmetry group is a subgroup of the symplectic diffeomorphisms
of the cotangent
bundle of the world-sheet, and the $\W_N$ geometry is obtained from a
non-linear truncation
of the $\W_\infty$ geometry. Quantum \W-gravity is briefly discussed.}

 \endpage
\pagenumber=1

\def\N {{\cal N }}

\def\np{Nucl. Phys.}
\def\pl{Phys. Lett.}

\def\cmp{Comm. Math. Phys.}
\def\intmp{Intern. J. Mod. Phys.}

\def\half{{\textstyle {1 \over 2}}}

\def\IR{\relax{\rm I\kern-.18em R}}

\def\weta{{ \tilde g}}
\def\dm {\partial_{\mu}}
\def\dn {\partial_{\nu}}
\def\dr {\partial_{\rho}}

\def\ix {\int\!\!d^2x\;}
\def\intt {\int\!\! }

\def\ep {\epsilon^{\mu\nu}}

\def\gmn {g_{\mu\nu}}

\def\dpl {\partial_+}
\def\dmi {\partial_-}

\def\dalemb#1#2{{\vbox{\hrule height .#2pt
        \hbox{\vrule width.#2pt height#1pt \kern#1pt
                \vrule width.#2pt}
        \hrule height.#2pt}}}
\def\square{\mathord{\dalemb{5.9}{6}\hbox{\hskip1pt}}}

\REF\pol{A.M. Polyakov,   Phys. Lett. {\bf 103B} (1981) 207,211.}
\REF\bow{P. Bouwknegt and K. Schoutens, CERN preprint CERN-TH.6583/92,
to appear in Physics Reports.}
\REF\hu{C.M. Hull, \pl\ {\bf 240B} (1990) 110.}
 \REF\van{K. Schoutens, A. Sevrin and
P. van Nieuwenhuizen, \pl\ {\bf 243B} (1990) 245.}
 \REF\pop{E. Bergshoeff, C.N. Pope, L.J. Romans, E.
Sezgin, X. Shen and K.S. Stelle, \pl\ {\bf 243B}
(1990) 350; E. Bergshoeff, C.N. Pope and K.S. Stelle, \pl\ (1990).}
\REF\huee{C.M. Hull, \pl\ {\bf 259B} (1991) 68.}
\REF\hue{C.M. Hull,  \np\ {\bf 353
B} (1991) 707.}
\REF\meeee{C.M. Hull, \np\ {\bf 364B}
(1991) 621;
 \pl\ {\bf 259B} (1991) 68.}
\REF\vann{K. Schoutens, A.
Sevrin and P. van Nieuwenhuizen, \np\ {\bf B349}
(1991) 791   and Phys.Lett. {\bf 251B} (1990) 355.}
\REF\mik{A. Mikovi\' c, \pl\ {\bf 260B} (1991) 75.}
\REF\mikoham {A. Mikovi\' c,
\pl\ {\bf 278B} (1991) 51.}
\REF\wrev{C.M. Hull, in {\it Strings and Symmetries 1991}, ed. by  N.
Berkovits et al,  World Scientific, Singapore Publishing, 1992.}
\REF\wanog{C.M. Hull, \np\ {\bf B367} (1991) 731.}
\REF\wstr{C.N. Pope, L.J. Romans and K.S. Stelle, \pl\ {\bf 268B}
(1991) 167 and {\bf 269B}
(1991) 287.}
\REF\hegeom{C.M. Hull, \pl\ {\bf 269B} (1991) 257.}
\REF\hegeoma{C.M. Hull, \W-Geometry, QMW preprint, QMW-92-6 (1992),
 hep-th/9211113.}
\REF\wnprep{C.M. Hull, \lq The Geometric Structure of $\W_N$ Gravity', in
preparation.}
\REF\sot{G. Sotkov and M. Stanishkov, \np\ {\bf B356} (1991) 439;
G. Sotkov, M. Stanishkov and C.J. Zhu, \np\ {\bf B356} (1991) 245.}
\REF\wot{A. Bilal, \pl\ {\bf 249B} (1990) 56;
A. Bilal, V.V. Fock and I.I. Kogan, \np\ {\bf B359} (1991) 635.}
\REF\wit{E. Witten, in {\it Proceedings of the Texas A\& M Superstring
Workshop,
 1990}, ed. by R. Arnowitt et al, World Scientific Publishing, Singapore,
1991.}
\REF\bers{ M. Berschadsky and H. Ooguri,
\cmp\ {\bf 126} (1989) 49.}
\REF\germat{J.-L. Gervais and Y. Matsuo, \pl\ {\bf B274} 309 (1992) and
Ecole Normale
preprint LPTENS-91-351 (1991); Y. Matsuo, \pl\ {\bf B277} 95 (1992)}
\REF\itz{P. Di Francesco, C. Itzykson and J.B. Zuber, Commun. Math. Phys.
{\bf 140} 543 (1991).}
\REF\ram{J.M. Figueroa-O'Farrill, S. Stanciu and E. Ramos, Leuven preprint
KUL-TF-92-34.}
\REF\zam{A.B. Zamolodchikov, Teor. Mat. Fiz. {\bf 65} (1985)
1205.}
\REF\fatty{V.A. Fateev and S. Lykanov, \intmp\ {\bf A3} (1988) 507.}
 \REF\bil{A. Bilal and J.-L. Gervais, \pl\ {\bf
206B} (1988) 412; \np\ {\bf B314} (1989) 646; \np\ {\bf B318} (1989) 579;
Ecole Normale preprint LPTENS 88/34.}
 \REF\winf{I. Bakas, \pl\ {\bf B228} (1989) 57.}
\REF\park{Q-Han Park, \pl\ {\bf 236B} (1990) 429, {\bf 238B}
(1990) 287 and {\bf 257B} (1991) 105.}
\REF\romans{L.J. Romans, \np\ {\bf B352} 829.}
 \REF\riemy{B. Riemann, (1892) {\it Uber die Hypothesen welche der Geometrie
zugrunde liegen}, Ges. Math. Werke, Leibzig, pp. 272-287.}
\REF\finre{H. Rund, {\it The Differential Geometry of Finsler Spaces}, Nauka,
Moscow, 1981; G.S. Asanov, {\it Finsler Geometry, Relativity and Gauge
Theories}, Reidel, Dordrecht, 1985.}
\REF\mon{T. Aubin, {\it Non-Linear Analysis on Manifolds. Monge-Ampere
Equations}, Springer Verlag, New York, 1982.}
 \REF\pleb{J.F. Plebanski, J. Math. Phys. {\bf 16} (1975) 2395.}
\REF\pen{R. Penrose, Gen. Rel. and Grav., {\bf 7} (1976) 31.}
\REF\lin{U. Lindstrom and M. Ro\v cek, \np {\bf B222} (1983) 285.}
\REF\hit{
N. Hitchin, A. Karlhede, U. Lindstrom and M. Ro\v cek, Commun. Math.
Phys. 108 (1987) 535. }
 \REF\strom{A.
Strominger, \cmp\ {\bf 133} (1990) 163.}
\REF\manspen{P. Mansfield and B. Spence, \pl\ {\bf B258} (1991) 49.}
\REF\vanlig{K. Schoutens, A. Sevrin and
P. van Nieuwenhuizen, \np\ {\bf B364} (1991) 584 and {\bf B371} (1992) 315;
H. Ooguri, K. Schoutens, A. Sevrin and
P. van Nieuwenhuizen, \cmp\ {\bf 145} (1992)  515.}

\chapter{Introduction}

Infinite-dimensional symmetry algebras
play a central r\^ ole in two-dimensional physics
and there is an intimate relationship between such algebras  and
two-dimensional
gauge theories or string theories.
Perhaps the most important example is the Virasoro algebra, which is a
symmetry algebra of any two dimensional conformal field theory.
This infinite-dimensional rigid symmetry can be promoted to a local symmetry
(two-dimensional diffeomorphisms) by coupling the two-dimensional field
theory to gravity, resulting in a theory that is Weyl-invariant as well as
diffeomorphism invariant. The two-dimensional metric  enters the theory as
a Lagrange multiplier imposing   constraints which satisfy the Virasoro
algebra, and the Virasoro algebra also emerges as the residual symmetry
that remains after choosing a conformal gauge.
 The quantisation of such a system of matter coupled to gravity
defines a   string theory and if the matter system is chosen such that
the world-sheet metric $h_{\mu \nu}$ decouples from the quantum theory
(\ie\ if the matter central charge is $c=26$), the string theory is said to
be critical [\pol]. Remarkably, the introduction of   gravity on the
world-sheet
leads to a critical string theory which leads to gravity in space-time.

The situation is similar for each of the cases in the following table.
\medskip
\smallskip
{\Tenpoint {
\def\thalf{{\textstyle {3 \over 2}}}
\settabs\+ &   $N=2$ Super-Virasoro ~   &  $2,\thalf,\thalf,1$   ~ &
  Topological Gravity  ~ &   $h_{\mu
\nu},\psi_\mu, \bar \psi_\mu, A_\mu$  ~ & \it $N=2$ Superstring
  \cr \+&\it Algebra &\it Spins  &\it 2-D Gauge Theory   &\it
Gauge Fields  & \it String Theory \cr
\smallskip
\+&Virasoro Algebra &2& Gravity &$h_{\mu \nu}$ &Bosonic String\cr
\+&Super-Virasoro  &$2,\thalf$& Supergravity &$h_{\mu \nu},\psi_\mu$
&Superstring\cr
\+&$N=2$ Super-Virasoro  &$2,\thalf,\thalf,1$&$N=2$ Supergravity &$h_{\mu
\nu},\psi_\mu, \bar \psi_\mu, A_\mu$ &$N=2$ Superstring\cr
\+&Topological Virasoro   &$2,2,1,1$& Topological Gravity &$h_{\mu
\nu},g_{\mu
\nu}, A_\mu,\psi_\mu$ &Topological String\cr
\+&\W-Algebra   &$2,3,\dots$&\W-Gravity &$h_{\mu
\nu},B_{\mu
\nu \rho}, \dots$ &\W-Strings\cr  }}
\smallskip
\medskip
\noindent In the first column are extended conformal algebras, \ie\ infinite
dimensional algebras that contain the Virasoro algebra. Each algebra is
generated by a set of currents, the spins of which are given in the second
column. Each algebra can arise as the symmetry algebra of a particular class
of conformal field theories e.g. the super-Virasoro algebra is a symmetry of
super-conformal field theories
while the topological Virasoro algebra is a
symmetry of topological
conformal field theories. For such
theories, the infinite-dimensional rigid
symmetry of the matter system
can be promoted to a local symmetry by coupling to the gauge theory
listed in the third column. In this coupling, the currents generating the
extended conformal algebra are coupled to the  corresponding gauge fields in
the fourth column. In each case, the gauge fields enter as Lagrange
multipliers and the constraints that they impose satisfy the algebra given in
the first column. Finally, integration over the matter and gauge fields
defines a generalisation of string theory which is listed in the last column.
In general, the gauge fields will become dynamical in the quantum theory, but
for special choices of conformal matter system (e.g. $c=26$ systems for the
bosonic string, $c=0$ systems for the topological string or $c=15$ systems
for the $N=1$ superstring), the string theory will be \lq critical' and the
gauge fields will decouple from the theory. A row can be added to the table
corresponding to any two-dimensional extended conformal algebra.

Consider now the set of models corresponding to the last row of
the table. A \W-algebra might be defined as any extended conformal algebra,
\ie\ a closed algebra that satisfies the Jacobi identities,   contains
the Virasoro algebra as a subalgebra and is generated by a (possibly
infinite) set of chiral currents (for a review, see [\bow]).
 Often the definition of \W-algebra
is restricted    to those algebras for which at least one of the generating
currents has spin greater than $2$, but relaxing this condition allows the
definition to include all the algebras in the  table and almost all of the
results to be reviewed here apply with this more general definition.
However,
many (but not all) interesting \W-algebras contain a spin-three
current, and for this reason $3$ is included as a typical higher spin
in the \W-algebra entry in the
  table.

The infinite-dimensional \W-algebra symmetry can be promoted to a local
symmetry
by coupling to suitable gauge fields, to obtain a theory of  matter coupled to
\W-gravity [\hu-\mikoham]; for a review, see [\wrev]. Again, the gauge fields
will in general become dynamical in the quantum theory [\wanog] but for
matter systems for which the \W-algebra central charge takes a particular
critical value, the \W-gravity gauge fields decouple and it is possible to
interpret the theory as a \W-string [\wstr]; in a \W-conformal gauge,
this can be thought of as a conformal field theory satisfying constraints
which generate a
\W-algebra instead of the usual Virasoro constraints.

The simplest \W-algebras are those that are Lie algebras, with the generators
$t_a$ (labelled by an index $a$ which will  in general have an infinite range)
satisfying commutation relations of the form
$$
[t_a,t_b]={f_{ab}}^ct_c +c_{ab}
\eqn\lie$$
for some structure constants ${f_{ab}}^c$ and constants $c_{ab}$, which define
a central extension of the algebra. However, for many \W-algebras, the
commutation relations give a result non-linear in the generators
$$
[t_a,t_b]={f_{ab}}^ct_c +c_{ab}+{g_{ab}}^{cd}t_ct_d+\dots =F_{ab}(t_c)
\eqn\nonlie$$
and the algebra can be said to close in the sense that the right-hand-side is a
function of the generators.
Most of the \W-algebras that are generated by a finite number of currents,
with at least one current of spin greater than two, are non-linear algebras of
this type.
Classical \W-algebras for which the bracket in \nonlie\ is a Poisson
bracket are straightforward to define, as the non-linear terms on the
right-hand-side can be taken to be a   product of classical
charges. For quantum \W-algebras, however,   the bracket is realised as a
commutator of quantum operators and the definition of the right-hand-side
requires some normal-ordering prescription. The complications associated
with the normal-ordering mean that there are classical \W-algebras for which
there is no corresponding quantum \W-algebra that satisfies the Jacobi
identities [\hue]. At first sight, it appears that there might be a problem in
trying to realise a non-linear algebra in a field theory, as symmetry
algebras are usually Lie algebras. However, as will be seen, a non-linear
algebra can be realised as a symmetry algebra for which the structure \lq
constants' are replaced by field-dependent quantities.

Consider a field theory in flat Minkowski space with metric $\eta_{\mu \nu}$
and coordinates $x^0,x^1$. The stress-energy tensor is a symmetric tensor
$T_{\mu \nu}$ which, for a translation-invariant theory, satisfies the
conservation law $$
\partial ^\mu
T_{\mu \nu}=0
\eqn\tcon$$
A spin-$s$ current in flat two-dimensionsal space is a rank-$s$ symmetric
tensor $W_{\mu_1 \mu_2 \dots \mu _s}$\foot{Recall that, in two dimensions, any
tensor can be decomposed into a set of symmetric tensors, e.g.
$V_{\mu \nu}=V_{(\mu \nu)}+V\epsilon _{\mu \nu}$ where $V=\half \epsilon ^{\mu
\nu}V_{\mu \nu}$. Thus without loss of generality, all the conserved
currents of a given theory can be taken to be symmetric tensors. A rank-$s$
symmetric tensor transforms as the spin-$s$ representation of the
two-dimensional Lorentz group.} and will be conserved if $$ \partial ^{\mu
_1} W_{\mu_1 \mu_2 \dots \mu _s}=0 \eqn\wcon$$ A theory is conformally
invariant if the stress tensor is traceless, ${T_\mu}^\mu=0$. Introducing
null coordinates $x^\pm={1 \over \sqrt 2} (x^0 \pm x^1)$, the tracelessness
condition becomes $T_{+-}=0$ and \tcon\ then implies that the remaining
components $T_{\pm\pm}$ satisfy
$$ \dpl T_{--}=0, \qquad \dmi T_{++}=0
\eqn \rert$$
If a spin-$s$ current $W_{\mu_1 \mu_2 \dots \mu _s}$ is traceless, it will
have only two non-vanishing components, $W_{++ \dots  +}$
and $W_{--\dots -}$. The conservation condition \wcon\ then
implies that
$$ \dmi W_{++ \dots  +}=0, \qquad \dpl W_{--\dots -}=0\eqn\dfg$$
so that $W_{++ \dots  +}=W_{++ \dots  +}(x^+)$
and $W_{--\dots -}=W_{--\dots -}(x^-)$ are right- and left-moving chiral
currents, respectively.
For a given conformal field theory, the set of all right-moving chiral
currents generate a closed current algebra, the right-moving chiral algebra,
and similarly for left-movers. The right and left chiral algebras are
examples of \W-algebras but are often too large to be useful. In studying
conformal field theories, it is often useful to restrict attention to all
theories whose chiral algebras contain a particular \W-algebra; the
representation theory of that \W-algebra then gives a great deal of useful
information about the spectrum, modular invariants etc of those theories
and may lead to a classification.

A field theory with action $S_0$ and symmetric tensor conserved currents
$T_{\mu \nu},W^A_{\mu_1 \mu_2 \dots \mu _{s_A}}$
(where $A=1,2,\dots $ labels the currents, which have spin $s_A$) will be
invariant under rigid symmetries  with constant parameters $k^\mu, \lambda
_A
^{\mu_1 \mu_2 \dots \mu _{s_A-1}}$ (translations and \lq
\W-translations') generated by the Noether charges $P_\mu,
 Q^A_{\mu_1 \mu_2 \dots \mu _{s_A-1}}$
(momentum and \lq \W-momentum') given by
  $P_\mu=\intt
dx^0T_{0\mu}$ and  $ Q^A_{\mu_1 \mu_2 \dots \mu _{s_A-1}}=\intt dx^0
W^A_{\mu_1 \mu_2 \dots \mu _{s_A-1}0}$. This is
true of non-conformal theories (e.g. affine Toda theories) as well as
conformal ones. However, if   the currents are traceless, then the theory
is in fact invariant under an infinite dimensional extended conformal
symmetry. The parameters $ \lambda
_A
^{\mu_1 \mu_2 \dots \mu _{s_A-1}}$
are then
 traceless and the corresponding transformations will be symmetries if the
parameters are not constant but satisfy the conditions that the
trace-free parts of $\partial ^{(\nu}k^{\mu)}, \partial ^{(\nu}
 \lambda
_A
^{\mu_1 \mu_2 \dots \mu _{s_A-1})}$ are zero.
This implies that
$\partial_{\mp}k^ \pm=0$ and $ \partial_{\mp}
\lambda _A^{\pm \pm \dots \pm} = 0$ so that
 the parameters are \lq semi-local', $k^\pm = k^\pm(x^\pm)$ and $\lambda
_A^{\pm \pm \dots \pm} =  \lambda _A^{\pm \pm \dots \pm}(x^\pm)$
and these are the parameters of conformal and \lq \W-conformal'
transformations.

The rigid symmetries corresponding to the currents $T_{\mu \nu},W^A_{\mu_1
\mu_2 \dots \mu _{s_A}}$ can be promoted to local ones by coupling to
the \W-gravity gauge fields
$h^{\mu \nu}, B_A^{\mu_1
\mu_2 \dots \mu _{s_A}}$ which are symmetric tensors transforming as
$$
\delta h^{\mu \nu}=\partial ^{(\nu}k^{\mu)}+\dots ,\qquad
\delta B_A^{\mu_1
\mu_2 \dots \mu _s}=\partial ^{(\nu}
\lambda_A ^{\mu_1 \mu_2 \dots \mu _{s-1})}+\dots,
\eqn\traaaa$$
to lowest order in the gauge fields. The action is given by the Noether
coupling
$$
S=S_0+ \ix \left( h^{\mu \nu}T_{\mu \nu}+ B_A^{\mu_1
\mu_2 \dots \mu _{s_A}}W^A_{\mu_1
\mu_2 \dots \mu _{s_A}}\right) + \dots
\eqn\snoth$$
plus terms non-linear in the gauge fields. If the currents
$T_{\mu \nu},W^A_{\mu_1
\mu_2 \dots \mu _s}$
are traceless, \ie\ if there is extended conformal symmetry, then the traces of
the gauge fields decouple and the classical
theory is invariant under Weyl and \lq
\W-Weyl' transformations given to lowest order in the gauge fields by
$$
\delta h^{\mu \nu}= \Omega \eta  ^{\mu \nu}+\dots ,
\qquad
\delta B_A^{\mu_1
\mu_2 \dots \mu _s}=\Omega_A ^{(\mu_1
\mu_2 \dots \mu _{s-2}}\eta  ^{\mu _{s-1}\mu_s)}+\dots
\eqn\wweyl$$
where $\Omega(x^\nu)$, $\Omega_A^{ \mu_1
\mu_2 \dots \mu _{s-2}}(x^\nu)$ are the local parameters.
This defines the linearised coupling to \W-gravity. The full non-linear
theory is in general non-polynomial in the gauge fields of spins 2 and higher.

The key to the non-polynomial structure of the coupling of matter to
gravity is the tensor calculus and the construction of the  fundamental
density $\sqrt g$. These have a natural interpretation in terms of
Riemannian geometry, which is based on a line element
$ds=(\gmn d x^\mu d x ^\nu)^{1/2}$.
A higher spin generalisation of Riemannian geometry is needed in order to
describe the geometry of \W-gravity. A number of approaches have been proposed
in
the literature [\sot-\ram];
here we shall review the approach of [\hegeom,\hegeoma,\wnprep].
 A spin-$n$ gauge field could be used
to define a geometry based on a line element
$
ds=(g_{\mu_1 \mu_2 \dots \mu _n} d x^{\mu_1} d x^{\mu_2}\dots d
x^{\mu_n})^{1/n} $ (first considered by Riemann [\riemy]).
A further generalization is to consider a line element
$
ds=N(x,dx)
 $
where $N$ is some function which is required to satisfy the homogeneity
condition
$
N(x,\lambda dx)= \lambda N(x,dx)
 $. This defines a Finsler geometry [\finre] and generalises the
spin-$n$ Riemannian line element.
To describe \W-gravity, it seems appropriate to generalise still further
and drop the homogeneity condition on $N$. Instead of working with
$N^2(x,dx)$, which is a function on the tangent bundle of the space-time
$M$ generalising the metric ($N^2=\gmn dx^\mu dx^\nu+...$), it will prove
more convenient to consider functions $F(x,y)$ on the cotangent bundle of
$M$,  generalising the inverse metric ($F(x^\mu , y_\mu)=
g^{\mu \nu } (x) y_\mu y_\nu +\dots$). Expanding in the fibre coordinate
$y$ gives a set of higher spin gauge fields (see eqn (6.1)).
Such a geometric interpretation of the results reviewed here seems
appropriate as they
 have a natural expression in terms of functions on the
cotangent bundle; for further discussion, see [\hegeoma].

Given a generalised line element $F(x^\mu , y_\mu)$, one can demand that it
be preserved under diffeomorphisms $ x \rightarrow f(x)$, and so derive the
tensorial transformation rules of the coefficients $g^{\mu \nu } (x) $, $
g^{\mu \nu \rho} (x) ,\dots$ occuring in the Taylor expansion of $F(x,y)$
with respect to $y$.
However, one can consider a much larger group of transformations
of the form $ x \rightarrow f(x,y)$, $ y \rightarrow g(x,y)$ consisting of
diffeomorphisms of the cotangent bundle $T^*M$ and it is natural to ask whether
such transformations could lead to the higher -spin symmetries of \W-gravity.
Expanding the infinitesimal transformation
$\delta x^\nu = k^\nu (x,y) = \sum k   (x)^{\nu\mu_1 ... \mu_n}
y_{\mu_1}...y_{\mu_n}$ gives a set of higher-spin transformations
($k ^{\nu\mu_1 ... \mu_n}$ is the parameter for a local spin-($n+2$)
transformation). However, this group of transformations turns out to be too
large
for the present purposes;  it is much larger than the symmetry group of
$w_\infty$-gravity. However, the subgroup of this consisting of symplectic
diffeomorphisms of $T^*M$ turns out to play an important role in \W-gravity
and contains the symmetry group of \W-gravity.

If $M$ is a $D$-dimensional space with coordinates $x^\mu$,
the cotangent bundle $T^*M$ is a
$2D$-dimensional space whose coordinates can be taken to be $x^\mu, y_\mu$
where $y_\mu$ transform as covariant vectors under diffeomorphisms of
$M$. The cotangent bundle is a symplectic manifold with natural two-form
$ \Omega=dx^{\mu}_{\Lambda}dy_{\mu}$ and
the subgroup of the diffeomorphisms of $T^*M$ that preserve
the symplectic structure $\Omega$ (\ie\ the canonical
transformations for the phase space $T^*M$ with coordinates $x^\mu$ and
momenta $y_\mu$) is the group of
 symplectic diffeomorphisms
of $T^*M$.

The symplectic diffeomorphisms have an interesting field theoretic
realisation. Let $\phi(x)$ be a real scalar field on $M$, which defines a
section
  $y_{\mu}(x)=\partial_{\mu}\phi$ of $T^*M$. Consider the transformation
$$\delta\phi=\sum^{\infty}_{n=2}
\lambda^{\mu_{1}\mu_{2}...\mu_{n-1}}_{(n)}(x^{\nu
})\partial_{\mu_{1}}\phi\partial_{\mu_{2}}\phi....\partial_{\mu_{n-1}}\phi
\equiv\Lambda(x^{\mu},y_{\mu})
\eqn\sdtrans$$
where
$y_{\mu}=\partial_{\mu}\phi
$
and the $\lambda^{\mu_{1}\mu_{2}...\mu_{n-1}}_{(n)}(x^{\nu
})$ ($n=2,3,\dots$) are infinitesimal parameters which are symmetric
tensor fields on $M$.
These transformations satisfy the algebra
$$\lbrack\delta_{\Lambda},\delta_{\Lambda^\prime}\rbrack=\delta_{\lbrace
\Lambda,\Lambda^\prime\rbrace}\eqn\alg$$
where the Poisson bracket is given  by
$$\lbrace\Lambda,\Lambda^\prime\rbrace={\partial\Lambda\over\partial
x^{\mu}}{\partial\Lambda^\prime\over\partial y_{\mu}}-{\partial\Lambda
^\prime\over\partial x^{\mu}}{\partial\Lambda\over\partial y_{\mu
}}
\eqn\poiss$$
(Note that the $\dm y_\nu$ terms cancel in \poiss.)
This is precisely the algebra of symplectic diffeomorphisms of $T^*M$.
In the one-dimensional case $M=S^1$,  $T^*M$ is the cylinder
$S^1 \times \IR$ and the algebra of symplectic diffeomorphisms is the algebra
$w_\infty$ introduced in [\winf], while  \sdtrans\ is   the realisation given
in [\pop].

The $n=2$ term in \sdtrans\ is  $\delta \phi = \lambda _{(2)}^\mu \dm
\phi$, which is an infinitesimal coordinate transformation on $M$
generated by the vector field  $\lambda _{(2)}^\mu$, while for higher $n$,
\sdtrans\ gives a set of non-linear higher spin generalisations of this.
The transformations \sdtrans\ have been suggested as the basis of
higher-spin generalisations of gravity [\wit,\pop], but unfortunately actions
that are invariant under transformations of the type \sdtrans\ only exist in
one space-time dimension, $D=1$ [\hue,\hegeoma]. Nevertheless, in
two-dimensions,
it is known that a $w_\infty$ gravity theory exists and an implicit
construction
of the action has been given [\pop] using the methods of [\van]. Just as the
coupling of $\phi$ to  gravity gives a theory with a spin-two gauge
invariance which, on going to conformal gauge, reduces to a residual
symmetry consisting of two copies of the Virasoro algebra (the algebra of
diffeomorphisms of $S^1$), the
coupling of $\phi$ to  $w_\infty$-gravity gives a theory with
gauge invariances of spins $2,3,4,\dots$ which, on going to a \W-conformal
gauge, reduce  to a residual symmetry consisting of two copies of the
$w_\infty$ algebra (the algebra of symplectic diffeomorphisms of $T^*S^1$).
The geometry
of two-dimensional $w_\infty$-gravity and the relation of its symmetry to
the symplectic diffeomorphisms \sdtrans\ will now be analysed
and the truncation to $\W_N$ gravity  described.

\chapter{Linearised $w_\infty$ Gravity}

Linearised $w_\infty$ gravity can be constructed by perturbing
about a flat two-dimensional space $M_0$ with metric
$ds^{2}=\mathop{\eta_{\mu\nu}dx^{\mu}dx^{\nu}=}\nolimits2dzd\bar
{z}$, where
 $z={1\over\sqrt{2}}\left(x^{1}+ix^{2}\right)$, $\bar{z}={1\over\sqrt
{2}}\left(x^{1}-ix^{2}\right)$ are complex coordinates if $M$ is Euclidean,
while, if $M$ is Lorentzian,
$z={1\over\sqrt{2}}\left(x^{1}+x^{2}\right)$, $ \bar{z}={1\over\sqrt
{2}}\left(x^{1}-x^{2}\right)$ are null real coordinates.
If a rank $s$ symmetric tensor $T_{\mu_{1}\mu_{2}...\mu_{s}}$ is traceless,
$\eta^{\mu\nu}T_{\mu\nu\rho \sigma...}=0$,
then it has only two
non-vanishing components, $ T_{zzz...z}$ and $ T_{\bar
{z}\bar{z}\bar{z}... \bar{z}}$.

Consider the free scalar field action
$$
S_0={1 \over 2} \ix   \dm \phi \partial ^\mu \phi
\eqn\free$$
This has an infinite number of conserved currents, which include [\pop]
$$W_{n}={1\over n}(\partial\phi)^{n} , \qquad n=2,3,....
\eqn\wis$$
 and these satisfy the conservation law $\mathop{\bar \partial
W}\nolimits_{n}=0$
 (where $ \partial=\partial_{z}$,
$\bar{\partial}=\partial _{\bar{z}}$). These currents generate the
transformations $$\delta_{L}\phi=\sum^{\infty}_{n=2}\lambda_{n}(z)(\partial
_{z}\phi)^{n-1}\equiv L(z,\partial_{z}\phi)
\eqn\wco$$
which are symmetries of \free\ provided the parameters are holomorphic,
$ \bar \partial\lambda_{n}=0 $. For $n=2$,
$  W_{2}=T_{zz}\  $ is the   stress tensor and the $\lambda _2$ term in
\wco\ is a   conformal transformation.
These transformations satisfy the $w_ \infty$ algebra
$\lbrack\delta_{L},\delta_{L^\prime}\rbrack=\delta_{\lbrace L,L^\prime
\rbrace}$ where, for any
 $L(z,w)$, $L^\prime(z,w)$ with $w=\partial_{z}\phi$, the Poisson bracket
is  $\lbrace L,L^\prime\rbrace={\partial _z L }{\partial _w L^\prime
 }-{\partial _zL^\prime }{\partial _w L} $.
Similarly, the anti-holomorphic currents
 $\mathop{\overline{W}}\nolimits_{n}={1\over n}(\bar \partial\phi)^{n}$
generate a second copy of $w_\infty$ (which commutes with the first)
and correspond to transformations
 $\delta \phi=\sum^{\infty}_{n=2}\bar \lambda_{n}(\bar z)(\bar \partial
 \phi)^{n-1}$.
Note that with a Lorentzian signature, $\lambda$ and $\bar \lambda$
are independent real parameters, but in Euclidean signature, they are
complex and satisfy
 $\mathop{\bar{\lambda}}\nolimits_{n}=(\lambda_{n})^{*}$.

The parameters $\lambda_{n}, \bar \lambda_{n}$ can be regarded as the two
components of a rank $n-1$ symmetric tensor
$\lambda^{\mu_{1}\mu_{2}...\mu_{n-1}}_{(n)}$
satisfying the tracelessness condition
$$\eta_{\mu\nu}\lambda^{\mu\nu\rho....\sigma}_{(n)}=0
\eqn\trlam$$
with
 $\lambda_{n}=\lambda^{zzz...z}_{(n)}$, $\mathop{\bar{\lambda}}\nolimits
_{n}=\lambda^{\bar{z}\bar{z}\bar{z}...\bar{z}}_{(n)} $.
The conditions $\bar \partial \lambda_{n}=0$, $ \partial \bar\lambda_{n}=0$
 become the condition that the trace-free part of $\partial ^{(\tau}
\lambda^{\mu\nu\rho....\sigma )}_{(n)}$ vanishes.
Then the $\lambda_{n}$ and $ \bar \lambda_{n}$ transformations
can be rewritten as \sdtrans\ with the parameters
 $\lambda^{\mu ....\sigma}_{(n)}$ satisfying these two constraints;
the constraint \trlam\ can be rewritten as
$$\eta_{\mu\nu}{\partial^{2}\Lambda\over\partial y_{\mu}\partial y_{\nu
}}=0
\eqn\linlam$$

Consider now the promotion of the \lq \W-conformal symmetry' with
(anti-) holomorphic parameters $\lambda_{n}(z)$ ($\bar \lambda_{n}(\bar
z)$) to a full gauge symmetry  with local parameters $\lambda_{n}(z, \bar
z)$, $\bar \lambda_{n}(z, \bar
z)$
under which $\phi$ transforms as
$$\delta \phi=\sum^{\infty}_{n=2}\left[ \lambda_{n}(z,\bar{z})(\partial
\phi)^{n-1}  +\mathop{\bar{\lambda}}\nolimits_{n}(z,\bar{z})(\bar \partial
\phi)^{n-1} \right]
\eqn\tre $$
To do this, it is necessary to introduce    gauge fields $h_n, \bar h_n$
transforming as
 $$\delta h_{n}=-2\bar{\partial}\lambda_{n}+O(h),\ \
\delta\mathop{\bar
{h}}\nolimits_{n}=-2\partial\mathop{\bar{\lambda}}\nolimits_{n}+O(h)
\eqn\linhtr$$
Then, the Noether coupling action
given by
 $$S=\int
d^{2}x\left\lbrack\bar{\mathop{\partial\phi\partial}\nolimits
}\phi+ \sum^{\infty}_{n=2}{1\over
n}\left(h_{n}(\partial\phi)^{n}+\mathop{\bar
{h}}\nolimits_{n}(\bar{\partial}\phi)^{n}\right)+O(h^{2})\right\rbrack
\eqn\noeth$$
is invariant, to lowest order in the gauge fields, under \tre,\linhtr.
The full gauge-invariant action [\pop,\vann] is non-polynomial in the gauge
fields, and the   gauge transformations \tre,\linhtr\ are also modified by
non-polynomial terms.

The action \noeth\ can be rewritten as
$$S=\int d^{2}x\left\lbrack {1\over2}\eta ^{\mu\nu
}\partial_{\mu}\phi\partial_{\nu}\phi+ \sum^{\infty}_{n=2}{1\over
n}\tilde h^{\mu_{1} ...\mu_{n}}_{(n)}\partial_{\mu_{1}}\phi
...\partial_{\mu_{n}}\phi+O(\tilde h^{2})\right\rbrack\equiv \int
d^{2}x\tilde{F}(x,y) \eqn\renoeth$$
where the $\tilde h^{\mu_{1}\mu_{2}...\mu_{n}}_{(n)}$ are symmetric tensor
gauge fields satisfying
 $$\eta_{\mu\nu}\tilde h^{\mu\nu\rho...\sigma}_{(n)}=0+O(h^{2})
\eqn\traa$$
at least to lowest order in the gauge fields, so that the only
non-vanishing components are
  $h_n =\tilde h^{zzz....z}_{(n)}$, $\mathop{\bar{h}}\nolimits_{n}=\tilde
h^{\bar {z}\bar{z}\bar{z}...\bar{z}}_{(n)} $.
The constraint \traa\ can
be rewritten, to lowest order,  as (using \renoeth)
$$\eta_{\mu\nu}{\partial^{2}\over\partial y_{\mu}\partial
y_{\nu}}\left[\tilde{F}- {1 \over 2}y_\mu y^\mu \right]=0+....
\eqn\low$$

Note that it is not strictly necessary to impose the trace condition
\traa. If it is not imposed, then the action \renoeth\ remains invariant
  to linearised order, provided the extra gauge fields corresponding
to the traces of $\tilde h^{\mu\nu\rho...\sigma}_{(n)}$ are inert under
the gauge transformations, at least to lowest order in the gauge fields.
However, this theory is reducible in the sense that it has more gauge
fields than symmetries, and can be consistently truncated to one with gauge
fields satisfying \traa.
If \traa\ is imposed, then the constraint can be solved in terms of
unconstrained gauge fields
$  h^{\mu ...\sigma}_{(n)}$ by $\tilde h^{\mu ...\sigma}_{(n)}=
  h^{\mu ...\sigma}_{(n)}- \ traces$,
while the constraint \trlam\ can be solved in terms of unconstrained
parameters $k_{(n)}^{\mu...}$
by $\lambda^{\mu ...\sigma}_{(n)}=
  k^{\mu ...\sigma}_{(n)}- \ traces$.
 The gauge fields $  h^{\mu ...\sigma}_{(n)}$ can be taken to transform as
$$\delta h^{\mu \nu \rho...\sigma}_{(n)} = \partial ^{(\mu} k _{(n)}
^{\nu \rho \dots \sigma)}+\eta^{(\mu \nu} \sigma _{(n)}^{\rho \dots
\sigma)}+O(h) \eqn\erter$$
where $\sigma _{(n)}^{\mu _1 ...\mu _{n-2}}(x)$ is the parameter for
local \W-Weyl transformations which correspond, to linearised order, to
shifts of the traces of the gauge fields $h_{(n)}$. For $n=2$,
$\eta ^{\mu \nu}+h_{(2)}^{\mu \nu}$ is the linearised inverse metric
tensor and $\sigma_{(2)}$ is the parameter of Weyl transformations.

\chapter{Linear and Non-Linear Gravity}

Before proceding to non-linear \W-gravity, it is useful to compare
with gravity. The linearised coupling of a scalar to gravity
is given by setting $h_{(n)}=0$ for $n>2$ in \renoeth,
and the full action is non-polynomial
in the gauge field  $\tilde h^{\mu \nu}$.
The field  $\phi$ transforms as
$$ \delta \phi = \lambda ^\mu \dm \phi\eqn\vagr$$
corresponding to the $n=2$ part of \tre.
The full action  must be quadratic in $\dm \phi$
on dimensional grounds and so   takes
the form
$$
S={1 \over 2} \ix   \tilde g^{\mu \nu} \dm \phi \dn \phi
\eqn\grac$$
for some $\tilde g^{\mu \nu}(x)$.
Comparing with the linearised theory, one finds that $\tilde
g^{\mu \nu}$ is a non-polynomial
function of  $\tilde h^{\mu \nu}$. However,
it is useful to forget about the original
variables $\tilde h^{\mu \nu}$ and
instead to study the non-linear theory
using the variable $\tilde g^{\mu \nu}(x)$, which has the virtue that it
appears linearly in the action  \grac. The action \grac\
is  diffeomorphism invariant if the field $\tilde
g^{\mu \nu }(x) $
transforms as a tensor density, \ie\ \grac\ is
invariant under \vagr\ and
$$
\delta \tilde g^{\mu \nu }=\lambda^\rho
\dr \tilde g^{\mu \nu }
-2\tilde g^{\rho (\mu   }\dr \lambda^{\nu)}+
\tilde g^{\mu \nu }\dr \lambda^\rho
\eqn\dens$$
This action can be used for any manifold $M$
 and makes no reference to any background metric.

This formulation is reducible, as it uses three gauge fields (the
components of $\tilde g^{\mu \nu}$) for two gauge symmetries.
An irreducible theory is obtained by imposing the constraint
$$ det \left(\tilde g^{\mu \nu}(x) \right) = \epsilon
\eqn\geco$$
(where $\epsilon =1$  for Euclidean
signature or $\epsilon= -1$ for Lorentzian signature).
 This   is
preserved under \dens\ and so can be consistently imposed, and reduces to
$\eta _{\mu \nu}\tilde h^{\mu \nu}=0 +....$ (the $n=2$ part of \traa)
in the linearised theory, where
  $\tilde g^{\mu \nu}=\eta^{\mu \nu}+\tilde h^{\mu
\nu}+O(h^2)$.
The constraint \geco\ can   be solved in terms of an unconstrained tensor
$g^{\mu \nu }$ by writing $\tilde g^{\mu \nu }=\sqrt {\epsilon g}g^{\mu \nu
}$ where $g=
\left[ det \left(
g^{\mu \nu} \right) \right]^{-1 } $. This solution
is invariant under the Weyl symmetry   $g^{\mu \nu}
\rightarrow \sigma g^{\mu \nu}$, so that $\tilde g^{\mu \nu}$ depends on only
two of the three components of $g^{\mu \nu}$,  the other component being pure
gauge.

\chapter{Non-Linear \W-Gravity}

The non-linear structure of $w_\infty$ gravity will now be presented,
following the approach of the previous section. The proof of these
results will be given elsewhere. The action is a non-polynomial
function of $\dm \phi$ and can be written as
 $$S=\int _M d^{2}x\tilde{F}(x,\partial\phi)
\eqn\sis$$
for some $\tilde{F}$, which has the following expansion in $y_\mu =\dm \phi$:
$$\tilde{F}(x,y)=\sum^{\infty}_{n=2}{1\over n} {\tilde{g}}
^{\mu_{1}\mu_{2}...\mu_{n}}_{(n)}\mathop{(x)y}\nolimits_{\mu_{1}}y_{\mu
_{2}}...y_{\mu_{n}}
\eqn\fis$$
where ${\tilde{g}}
^{\mu_{1}\mu_{2}...\mu_{n}}_{(n)}(x)$ are gauge fields.
The action will be invariant under diffeomorphisms of $M$ if these gauge
fields are tensor densities on $M$.
No constraints will be imposed on the $\tilde g_{(n)}$ to start with, so
that the linearised form of this action is the reducible theory
given by \renoeth\ without the constraint \traa.
The next step is to  seek symmetries of this action whose linearised form
agrees with those of section 2. The variation of $\phi$ can be taken to be
 $$\delta\phi=\Lambda(x,\partial\phi)
\eqn\trfi$$
where
$$\Lambda(x^{\mu},y_{\mu})=\sum^{\infty}_{n=2}\lambda^{\mu_{1}\mu
_{2}...\mu_{n-1}}_{(n)}\mathop{(x)y}\nolimits_{\mu_{1}}y_{\mu_{2}}...y_{\mu
_{n-1}}
\eqn\liss$$
for some parameters $\lambda^{\mu_{1}\mu
_{2}...\mu_{n-1}}_{(n)}(x)$. From the linearised analysis, one would
expect there to be a symmetry of this kind only if a constraint is
imposed on the parameters which is given by \linlam\ plus higher order terms
in the gauge fields.
This is indeed the case and the full non-linear
constraint
  is $$
\epsilon ^{\mu \rho} \epsilon ^{\nu \sigma} {\partial ^2 \Lambda \over
\partial y_\mu \partial y_ \nu} {\partial ^2 \tilde F \over \partial y_\rho
\partial y_ \sigma}=0
\eqn\conl$$
where $\epsilon ^{\mu \nu}$ is the alternating tensor density on $M$.
This involves no background metric and expanding in $y$ gives a sequence
of non-linear algebraic constraints on the parameters which, in the
linearised approximation, reduce to \trlam.
With this constraint on the parameters, the action \sis\ is invariant (up
to a surface term) under the transformations given by \trfi,\liss\ and
$$\eqalign{
\delta\mathop{\tilde{g}}\nolimits^{\mu_{1}\mu_{2}...\mu_{p}}_{(p)}&=\sum
_{m,n=2}^\infty
\delta_{m+n,p+2}\biggl[(m-1)\lambda^{(\mu_{1}\mu_{2}...}_{(m)}\partial
_{\nu}\mathop{\tilde{g}}\nolimits^{...\mu_{p})\nu}_{(n)}-(n-1)\mathop{\tilde
{g}}\nolimits^{\nu(\mu_{1}\mu_{2}...}_{(n)}\partial_{\nu}\lambda^{...\mu
_{p})}_{(m)} \cr &
+{(m-1)(n-1)\over p-1}\partial_{\nu}\left\lbrace\lambda^{\nu(\mu
_{1}\mu_{2}...}_{(m)}\mathop{\tilde{g}}\nolimits^{...\mu_{p})}_{(n)}-\mathop{\tilde
{g}}\nolimits^{\nu(\mu_{1}\mu_{2}...}_{(n)}\lambda^{...\mu_{p})}_{(m)}\right
\rbrace\biggr]
\cr}
\eqn\denvar
$$
{}From \denvar, the $\tilde g_{(s)}$ transform as tensor
densities under the $\lambda _{(2)}$ transformations, as expected.

This gives an invariant action which is, however, reducible. To obtain the
irreducible theory, it is necessary to find constraints on the gauge
fields that are preserved by the transformations \denvar\ and which
reduce to \low\ in the linearised limit.
The unique such  constraint is given by
$$
det \left ({\partial ^2   \tilde F (x,y)\over \partial y_\mu
\partial y_ \nu} \right)=\epsilon
\eqn\decon$$
where
$\epsilon=1$ for Euclidean signature and $\epsilon=-1$ for Lorentzian
signature. This constraint takes a strikingly simple form which does not
depend on any background geometry.
 Expanding   \decon\ in $y_\mu$
gives an infinite number of constraints on the density gauge fields
$\tilde g_{(n)}^{\mu \nu \dots}$ (including \geco\ for $n=2$) which are
invariant under the transformations \denvar. If $\tilde F$ satisfies the
constraint \decon, the constraint \conl\ on the infinitesimal parameters
$\Lambda$ can be rewritten, to lowest order in $\Lambda$,
 as
$$
det \left ({\partial ^2   \over \partial y_\mu
\partial y_ \nu} [\tilde F +\Lambda](x,y) \right)=\epsilon
\eqn\lamcon$$

\chapter{The Non-Linear Constraints and Hyperkahler Geometry}

The constraint \decon\ is an equation of Monge-Ampere type [\mon] and
this leads to the following geometrical interpretation. Let $\zeta_\mu,
\bar \zeta _{\bar\mu}$
($\mu =1,2$) be complex coordinates on $\IR ^4$.
Then, for each $x^\mu$, a solution $\tilde F(x,y)$ of \decon\ can be used to
define a function $K_x(\zeta , \bar \zeta)$ on $\IR ^4$ by
$$K_x(\zeta , \bar \zeta)= \tilde F(x^\mu, \zeta_\mu +\bar \zeta _\mu)
\eqn\kis$$
For each $x$, $K_x$ can be viewed as the Kahler potential for a Kahler
metric on $\IR ^4$. As a result of \decon, each $K_x$ satisfies the
complex Monge-Ampere equation, sometimes referred to as the Plebanski
equation [\pleb], $det
(\partial _{ \mu} \partial _{ \bar \mu}K_x)=\epsilon$ and so the corresponding
metric is Kahler and Ricci-flat, which implies that the curvature tensor is
either
self-dual or anti-self-dual. In the Euclidean case, the metric has signature
$(4,0)$ and is hyperkahler, while in the other case the metric has signature
$(2,2)$ and holonomy $SU(1,1)$. As the Kahler potential is independent of
the imaginary part of $\zeta _\mu$, the metric has two
commuting (triholomorphic) Killing vectors, given by $i(\partial / \partial
\zeta _\mu - \partial / \partial \bar \zeta _{\bar \mu})$.
Thus the lagrangian $\tilde F(x,y)$ corresponds to a two-parameter family
of Kahler potentials $K_{x^\mu}$ for (anti) self dual geometries on $\IR ^4$
with two Killing vectors.
The parameter constraint \lamcon\
  implies that $\tilde F +\Lambda$ is  also   a Kahler
potential for a hyperkahler metric with two killing vectors, so that
for each $x$,  $\Lambda$ represents an infinitesimal deformation of the
hyperkahler geometry.

Techniques for solving the Monge-Ampere equation can be used to solve
\decon. The general solution of the Monge-Ampere equation  can be given
implicitly
 by   Penrose's twistor transform construction [\pen]. For solutions with one
(triholomorphic) Killing vector, the Penrose transform reduces to a
Legendre transform solution [\hit]  which was  found first in the context of
supersymmetric non-linear sigma-models [\lin]. Writing
 $y_{1}=\zeta$, $ y_{2}=\xi$,  any function
$\tilde{F}(x^\mu,\zeta,\xi)$ can be written as the Legendre transform with
respect to $\zeta$ of some $H$, so that
$$\tilde{F}(x,\zeta,\xi)=\pi\zeta -H(x,\pi,\xi)
\eqn\leg$$
where the equation
$${\partial H\over\partial\pi}=\zeta\ \eqn\treled$$
gives $\pi$ implicitly as a function of $x,\zeta,\xi$.
The  Monge-Ampere equation \decon\ will be satisfied if and only if $H$
satisfies the Laplace equation [\hit]
 $${\partial^{2}H\over\partial\pi^{2}}+\epsilon {\partial^{2}H\over\partial
\xi^{2}}=0
\eqn\htyhj$$
and the general solution of this is
$$H=f(x,\pi+\sqrt{- \epsilon} \xi)\ +\bar f(x,\pi-\sqrt {- \epsilon} \xi)
\eqn\hiss$$
where $f,\bar f$ are
  arbitrary independent real functions if $\epsilon =-1$ and are complex
conjugate functions if $\epsilon =1$. Then the general solution of \decon\
is  the Legendre transform \leg,\treled,\hiss\   and
  the action   can be given in the first order form
$$S=\ix\tilde{F}(x,y)=\ix \left(\pi \partial _\tau \phi -
 f(x^{\mu},\pi
+\partial _ \sigma \phi )+\bar f(x^{\mu},\pi-\partial _ \sigma \phi)
 \right)
\eqn\firs$$
where $\tau = x^1$ and $\sigma = -\sqrt{- \epsilon} x^2$.
The field equation for the auxiliary field $\pi$ is \treled\ and this can be
used in principle to eliminate $\pi$ from the action, but it will not be
possible to solve the equation \treled\   explicitly   in general. The
constraints \lamcon\
can be solved similarly.
 Expanding
the functions $f,\bar f$ gives the Hamiltonian form of the
$w_\infty$ action [\mik]
$$
S=\ix \left(
\pi \partial _\tau \phi- \sum _{n=2}^\infty {1 \over n}
\left[
h_n(\pi
+\partial _ \sigma \phi)^n +\bar h_n(\pi
-\partial _ \sigma \phi)^n
\right] \right)
\eqn\erertsf$$
consisting of a free term plus a set of Lagrange multiplier gauge fields,
$h_n,\bar h_n$,  times
  constraints. The symmetries of the action are simply those generated by
these constraints [\mik] (e.g. $\delta \phi
 =L(x^{\mu},\pi
+\partial _ \sigma \phi )+\bar L(x^{\mu},\pi
-\partial _ \sigma \phi )$ etc).

A generalisation of the Legendre transform solution that involves
transforming with respect to both components of $y_\mu$ and maintains
Lorentz covariance is suggested by the results of [\van,\pop]. Any
 $\tilde F(x,y)$ can be written
as a transform of a function $H$ as follows:
 $$\tilde{F}(
x^{\mu},y_{\nu})=2\pi^{\mu}y_{\mu}-{1\over2}\eta^{\mu
\nu}y_{\mu}y_{\nu}-2H(x,\pi)
\eqn\retywt$$
where the equation
$$y_{\mu}={\partial H\over\partial\pi^{\mu}}
\eqn\rtysd$$
implicitly determines
$\pi_{\mu}=\pi_{\mu}(x^{\nu},y_{\rho})$.
Then $\tilde F$ will satisfy \decon\ if and only if its transform $H$
satisfies
$${1 \over 2}
\eta^{\mu\nu}{\partial^{2}H\over\partial\pi_{\mu}\partial\pi_{\nu
}}= {\partial^{2}H\over\partial\pi_{+}\partial\pi_{-}}=1 \eqn\tyghas$$
The general solution of this is (with $\pi _\pm = \pi _1 \pm
\sqrt{-\epsilon} \pi _2$)
$$H=\pi_{+}\pi_{-}+f(x,\pi_{+})+\bar{f}(x,\pi_{-})\eqn\rtyzzeg$$ This
solution can be used to write the action
$$
S=
\int d^{2}x\left(2\pi^{\mu}y_{\mu}-\eta
_{\mu\nu}\pi^{\mu}\pi^{\nu}-{1\over2}\eta^{\mu\nu}y_{\mu}y_{\nu}-2f(x,\pi
_{+})-2\bar{f}(x,\pi_{-})\right)
\eqn\grhjgdjkgs$$
The field equation for $\pi^\mu $
is \rtysd, and using this to substitute for
$\pi$ gives the action \sis\ subject to the constraint \decon.
Alternatively, expanding the functions $f,\bar f$ as
$
f = \sum  {  s}^{-1}h_s(x)(\pi _+)^s
$, $
\bar f = \sum  {  s}^{-1}\bar h_s(x)(\pi _-)^s
 $
gives precisely the form of the action given in [\pop]. The parameter
constraint \lamcon\ is solved similarly, and the solutions can
be used to write the symmetries of \grhjgdjkgs\ in the form given in [\pop].

\chapter{Covariant Formulation and \W-Weyl Invariance}

In this section, the covariant
solution of the constraints given in [\hegeom]
will be reviewed. In [\hegeoma], an alternative covariant solution is given
which appears to have a deeper geometrical significance, but it would
take to long to descrie that here.
 The constraint \decon\ can be solved in terms of an unconstrained function
$$ {F}(x,y)=\sum^{\infty}_{n=2}{1\over n} { {g}}
^{\mu_{1}\mu_{2}...\mu_{n}}_{(n)}\mathop{(x)y}\nolimits_{\mu_{1}}y_{\mu
_{2}}...y_{\mu_{n}}
\eqn\fist$$
by writing
$$\tilde F(x,y)= \Omega (x,y) F(x,y)
\eqn\rerer$$
where
$\Omega $ is to be found in terms of $F$ and has an expansion of the form
$$ {\Omega}(x,y)=\sum^{\infty}_{n=0} { \Omega}
^{\mu_{1}\mu_{2}...\mu_{n}}_{(n+2)}\mathop{(x)y}\nolimits_{\mu_{1}}y_{\mu
_{2}}...y_{\mu_{n}}
\eqn\omis$$
Substituting \rerer\ in \decon\ gives a set of equations which can be solved
to give the tensors $\Omega_{(n)}$ in terms of the unconstrained tensors
$g_{(n)}$ in \fist, giving $\Omega _{(2)}= \sqrt {\epsilon g}$,
$\Omega _{(3)}^\mu= -\sqrt {\epsilon g}g_{(3)}^{\mu \nu \rho}g_{\nu \rho}$,
etc
where $g_{\mu \nu}$ is the inverse of $g^{\mu \nu}_{(2)}$ and $g=det
[g_{\mu \nu}]$. Substituting in \rerer\ gives
$$\tilde g^{\mu \nu}_{(2)}=\sqrt {\epsilon g}g^{\mu \nu}_{(2)},
\quad \tilde g_{(3)}^{\mu \nu \rho}=\sqrt {\epsilon g}\left[
g_{(3)}^{\mu \nu \rho} -{3 \over 2}g^{(\mu \nu}_{(2)}g_{(3)}^{  \rho)
\alpha \beta}
g_{\alpha \beta} \right], \dots
\eqn\yuouioio$$
Writing $\tilde F$ in terms of $F$ gives an action which is invariant
under the \W-Weyl transformations
$$\delta F(x,y)=\sigma(x,y)F(x,y)
\eqn\wweylf$$
Expanding
$$\sigma(x,y)=\sigma_{(2)}(x)+\sigma^{\mu}_{(3)}(x)y_{\mu}+\sigma
^{\mu\nu}_{(4)}(x)y_{\mu}y_{\nu}+...
\eqn\exsig$$
these can be written as
$$  \delta
g^{\mu\nu}_{(2)}=\sigma_{(2)}g^{\mu\nu}_{(2)}, \qquad \delta
g^{\mu\nu\rho}_{(3)}=\sigma_{(2)}g^{\mu\nu\rho}_{(3)}+{3 \over 2}\sigma
^{(\mu}_{(3)}g^{\nu\rho)}_{(2)},
\dots
\eqn\exweyl$$
These transformations can be used to remove all traces from the gauge
fields, leaving only   traceless gauge fields, as in [\pop,\vann].

The constraint \lamcon\ on the parameters $\lambda _{(n)}$ can be solved
in a similar fashion in terms of unconstrained parameters
$k_{(n)}^{\mu_1
...\mu_{n-1}}$ and the transformations of the unconstrained gauge fields
can be defined to take the form  $\delta { {g}}
^{\mu_{1}\mu_{2}...\mu_{n}}_{(n)}=\partial ^{(\mu_{1}}{ {k}}
^{\mu_{2}...\mu_{n})}_{(n)}+\dots$ (cf \erter). The $g_{(n)}$
might be thought of
as gauge fields for the whole of the symplectic diffeomorphisms of $T^*M$
(with parameters $k_{(n)}$), and appear in the action only through the
combinations $\tilde g_{(n)}$. The transformations of
$\tilde g_{(n)}$ and $\phi$ then only depend on the parameters
$k_{(n)}$ in the form $\lambda _{(n)}$.

\chapter{$\W _N$-Gravity}

A free scalar field in two dimensions
has the set of conserved currents \wis, given by
$W_{n}={1\over n}(\partial\phi)^{n} $, $n=2,3,....$,
and these generate a $w_{\infty}$ algebra.
In fact, the  finite subset of these given by $W_n$, $n=2,3,....,N$
generate a closed non-linear algebra which is a classical
limit of the $\W_N$ algebra, and in the limit $N \rightarrow \infty$, the
classical
current algebra becomes the $w_\infty$ algebra.
Similarly, the   currents
 $\mathop{\overline{W}}\nolimits_{n}={1\over n}(\bar \partial\phi)^{n}$
generate a second copy of the $\W_N$ or $w_\infty$ algebra.

The linearised action for $\W_N$ gravity is  given by
simply truncating the action \noeth\
by setting the gauge fields $h_n, \bar h_n$ with $n>N$ to zero, giving
$$S=\int
d^{2}x\left\lbrack  \partial\phi \bar \partial \phi+ \sum^{N}_{n=2}{1\over
n}\left[h_{n}(\partial\phi)^{n}+\mathop{\bar
{h}}\nolimits_{n}(\bar{\partial}\phi)^{n}\right]+O(h^{2})\right\rbrack
\eqn\noethn$$
which is invariant, to lowest order in the gauge fields, under the
transformations
 $$\eqalign{
\delta \phi&=\sum^{N}_{n=2}\left[
\lambda_{n}(z,\bar{z})(\partial \phi)^{n-1}
+\mathop{\bar{\lambda}}\nolimits_{n}(z,\bar{z})(\bar \partial \phi)^{n-1}
\right] \cr
\delta h_{n}&=-2\bar{\partial}\lambda_{n}+O(h),\qquad \
\delta\mathop{\bar
{h}}\nolimits_{n}=-2\partial\mathop{\bar{\lambda}}\nolimits_{n}+O(h)
\cr}
\eqn\tre$$
This gives the linearised action and transformations of $\W_N$ or
(in the $N\rightarrow \infty$ limit) $w_\infty$ gravity. The full
gauge-invariant action and gauge transformations  are non-polynomial in the
gauge fields.

The linearised action for
$\W_N$ gravity is then an $N$'th order polynomial in $\partial _\mu \phi$.
However, the full
non-linear action is non-polynomial in $\partial _\mu \phi$ and the gauge
fields $h_n$, but the
coefficient of $(\partial   \phi)^n$ for $n>N$ can   be written as a
non-linear function of the finite number of fundamental gauge fields
$h_2,h_3,\dots , h_N$  that
occur in the linearised action. The simplest way in which this might come
about would be if the action   were given by \sis,\fis\ and $\tilde F$
satisfies a constraint of the form
$${\partial^{N+1}\tilde{F}\over\partial y_{\mu
_{1}}\partial y_{\mu_{2}}...\partial y_{\mu_{N+1}}}=0+O(\tilde F^2)
\eqn\noo$$
where the right hand side is non-linear in  $\tilde F$ and its derivatives,
and depends only on derivatives of $ \tilde F$ of order $N$ or less.
This is indeed the case; the action for
$\W_N$ gravity is given by \sis\ where
$\tilde F$ satisfies \decon\ and \noo, and the right hand
side of \noo\ can be
given explicitly. Just as the non-linear constraint \decon\ had an
interesting
geometric interpretation, it might be expected that the non-linear form
of \noo\ should also be of geometric interest. Here, the
results will be summarised;
full details will be given in [\wnprep].

It will be useful to define
$$F^{\mu
_{1}\mu_{2}....\mu_{n}}(x,y)={\partial^{n}\tilde{F}\over\partial y_{\mu
_{1}}\partial y_{\mu_{2}}...\partial y_{\mu_{n}}}
\eqn\iuglg$$
and
$$H_{\mu\nu}(x,y)=2\left(\weta^{\mu\nu}+F^{\mu\nu}\right)^{-1}
\eqn\metis$$
where $\weta ^{\mu\nu}=\weta_{(2)}^{\mu\nu}(x)$.

The action for $\W_N$ gravity is then given by the action for $w_\infty$
gravity, but with the function $\tilde F$ satisfying one extra constraint
of the form \noo.
For $\W_3$, this extra constraint is
$$F^{\mu\nu\rho\sigma}={3\over2}H_{\alpha\beta}F^{\alpha(\mu\nu}F^{\rho
\sigma)\beta}
\eqn\eryh$$
or, using \metis,
$$F^{\mu\nu\rho\sigma}=3\left(\weta^{\alpha\beta}+
F^{\alpha\beta}\right)^{-1}F^{\alpha
(\mu\nu}F^{\rho\sigma)\beta}
\eqn\thcon$$
This is the required extra constraint
for $\W_3$ gravity.
Thus the action  for $\W_3$ gravity is given by \sis,\fis,
where $\tilde F$ is
a function satisfying the two constraints \decon\ and \thcon.

For $\W_4$ gravity, the extra constraint is
$$F^{\mu\nu\rho\sigma\tau}=5H_{\alpha\beta}F^{\alpha(\mu\nu}F^{\rho
\sigma\tau)\beta}-{15\over4}H_{\alpha\beta}H_{\gamma\delta}F^{\alpha
(\mu\nu}F^{\rho\sigma|\gamma|}F^{\tau)\beta\delta}
\eqn\fcon$$
so that the $\W_4$ action is \sis\ where $\tilde F$   satisfies
  \decon\ and \fcon, and $H^{\mu\nu}$ is given in
  terms of $\tilde F$ by \metis.
Similar results hold for all $N$. In each case, one obtains an equation of
the form  \noo, where the right hand side is constructed from the $n$'th
order derivatives $F^{\mu_1 \dots \mu_n}$ for $2<n\le N$ and from $H_{\mu
\nu}$.

Expanding $\tilde F$ in $\dm \phi$ \fis\ gives the coefficient of the
$n$-th order $\partial
_{\mu _1} \phi \dots \partial
_{\mu _n} \phi $ interaction, which is proportional to $\tilde g^
{\mu _1  \dots  \mu _n}_{(n)}$. The constraint \noo\ implies that for $n>N$,
the coefficient $\tilde g _{(n)}$ of the $n$-th order interaction
can be written in terms of the coefficients $\tilde g _{(m)}$
of the $m$-th order interactions for $2 \le m \le N$.  For $\W_3$, the
$n$-point vertex can be written in terms of 3-point vertices for $n>3$, so
that (with $\tilde{g}_{\alpha\beta}=\left(
\mathop{\tilde{g}}\nolimits_{(2)}^{\alpha\beta}\right
)^{-1}$)
$$\mathop{\tilde{g}}\nolimits_{(4)}^{\mu\nu\rho\sigma}=
\tilde{g}_{\alpha\beta}
\mathop{\tilde{g}}\nolimits_{(3)}^{\alpha(\mu\nu}\mathop{\tilde
{g}}\nolimits_{(3)}^{\rho\sigma)\beta}
\eqn\fion$$
$$\mathop{\tilde{g}}\nolimits_{(5)}^{\mu\nu\rho\sigma\tau}=
{5\over 4}
\tilde{g}_{\alpha\beta
}
\tilde{g}_{\gamma\delta}
\mathop{\tilde{g}}\nolimits_{(3)}^{\alpha
(\mu\nu}\mathop{\tilde{g}}\nolimits_{(3)}^{\rho\sigma|\gamma|}
\mathop{\tilde
{g}}\nolimits_{(3)}^{\tau)\beta\delta}
\eqn\fitw$$
etc, while for $\W_4$, all vertices can be written in terms
of 3- and 4-point
vertices, e.g.
$$\eqalign{
\mathop{\tilde{g}}\nolimits_{(5)}^{\mu\nu\rho\sigma\tau}
=&{5\over 2}
\tilde{g}_{\alpha\beta
}
\mathop{\tilde{g}}\nolimits_{(3)}^{\alpha(\mu\nu}\mathop{\tilde
{g}}\nolimits_{(4)}^{\rho\sigma\tau)\beta}
\cr &
-{5\over 4}
\tilde{g}_{\alpha\beta}
\tilde{g}_{\gamma\delta
}
\mathop{\tilde{g}}\nolimits_{(3)}^{\alpha(\mu\nu}\mathop{\tilde
{g}}\nolimits_{(3)}^{\rho\sigma|\gamma|}\mathop{\tilde{g}}\nolimits
_{(3)}^{\tau)\beta\delta}
\cr}
\eqn\fith$$
For the derivation of these results, and the form of the transformation
rules, see [\wnprep].

To attempt a geometric formulation of these results,
note that while the second derivative of $\tilde F$ defines a metric, the
fourth derivative is related to a   curvature, and the
$n$'th derivative is  related to the $(n-4)$'th covariant derivative of
the curvature. The $\W_3$ constraint \thcon\ can then be written
as  a constraint on the
curvature, while the  $\W_N$ constraint \noo\ becomes a constraint on the
$(N-3)$'th covariant derivative of
the curvature. One approach
is to
introduce a second Kahler metric $\mathop{\hat{K}}\nolimits_{x}$
on $\IR ^4$ given
in terms of the potential $K_{x}$ introduced in \kis\ by
$$\mathop{\hat{K}}\nolimits_{x}=K_{x}+\weta^{\alpha\bar{\beta}}\zeta
_{\alpha}\mathop{\bar{\zeta}}\nolimits_{\bar{\beta}}
\eqn\erter$$
The corresponding metric is given by
$$\mathop{\hat{G}}\nolimits^{\mu\bar{\nu}}=\weta^{\mu
\bar{\nu}} +G^{\mu\bar{\nu}}
\eqn\grfe$$
Then if $\tilde F$ satisfies the $\W_3$ constraint \thcon, the
curvature tensor for the metric \grfe\ satisfies
$$\mathop{\hat{R}}\nolimits^{\mu\bar{\nu}\rho\bar{\sigma}}={1\over
2}\mathop{\hat{G}}\nolimits_{\alpha\bar{\beta}}\left\lbrack
T^{\alpha
\mu\bar{\nu}}T^{\bar{\beta}\bar{\sigma}\rho}+T^{\alpha\mu\bar{\sigma
}}T^{\bar{\beta}\bar{\nu}\rho}+T^{\bar{\beta}\bar{\nu}\mu}T^{\alpha
\rho\bar{\sigma}}+T^{\bar{\beta}\bar{\sigma}\mu}T^{\alpha \rho \bar{\nu
}}\right\rbrack
\eqn\cur$$
where
$$T^{\mu\nu\bar{\rho}}={\partial^{3}\hat{K}\over\partial\zeta_{\mu
}\partial\zeta_{\nu}\partial\mathop{\bar{\zeta}}\nolimits_{\bar{\rho
}}},\ \ \ \ T^{\bar{\mu}\bar{\nu}\rho}={\partial^{3}\hat{K}\over\partial
\mathop{\bar{\zeta}}\nolimits_{\bar{\mu}}\partial\mathop{\bar{\zeta
}}\nolimits_{\bar{\nu}}\partial\zeta_{\rho}}
\eqn\tis$$
This is similar to, but distinct from, the constraint of special geometry
[\strom]. Note that \cur\ is not a covariant equation as the
definitions \tis\
are only valid in the special coordinate system that arises in the study of
\W-gravity. However,
  tensor fields $T^{\mu\nu\bar{\rho}},T^{\bar{\mu}\bar{\nu}\rho}$
  can be defined by requiring them to be
  given by \tis\ in the special coordinate system and to transform
  covariantly, in which case the equation \cur\ becomes
  covariant, as in the case of
   special geometry  [\strom].
For $\W_N$, this generalises to give a constraint on the
$(N-3)$'th covariant
derivative of the curvature, which is given in terms of
tensors that can each
be written in terms of  some higher order derivatives  of
the Kahler potential
in the special coordinate system.

For each $x^\mu$, the solutions to the constraints for $\W_N$ gravity are
parameterised by the $2(N-1)$ variables $h_n, \bar h_n$ for $2 \le n \le N$
which are then the coordinates for the
$2(N-1)$ dimensional
moduli space
for the  self-dual geometry satisfying the $\W_N$ constraint.
For the $x$-dependent family of solutions, the moduli become the
fields $h_n (x), \bar h_n(x)$ on the world-sheet.

\chapter{Conclusion}

We have seen that symplectic diffeomorphisms of the cotangent bundle of the
two-dimensional space-time (or world-sheet) $\N$ play a fundamental role in
\W-gravity, generalising the role played by the diffeomorphisms of $\N$ in
ordinary
gravity theories. Further, in the case in which the matter
system consists of
a single boson, we have completely determined the non-linear
structure of the
coupling to \W-gravity and found that it involves the
solution to an interesting
non-linear differential equation which can be linearised by a twistor
transform.
We found an infinite  set of
symmetric tensor density gauge fields
$\tilde g_{(n)}^{\mu_{1} ...\mu_{n}}$, $n=2,3\dots $,
 transforming under the action of a
gauge group isomorphic to the symplectic diffeomorphisms of
$ T^*\N $. The results could be simply stated in terms of the generating
function
$$\tilde{F}(x,y)=\sum^{\infty}_{n=2}{1\over n} {\tilde{g}}
^{\mu_{1}\mu_{2}...\mu_{n}}_{(n)}\mathop{(x)y}\nolimits_{\mu_{1}}y_{\mu
_{2}}...y_{\mu_{n}}
\eqn\fisab$$
and in [\hegeoma] it is argued that the quantity $\tilde F$
has a natural geometrical interpretation as a \lq \W-density'.
In [\hegeoma], the concept of a \lq \W-scalar' $F(x,y)$ was also introduced
which generated a set of tensor gauge fields
$ g_{(n)}^{\mu_{1} ...\mu_{n}}$, $n=2,3\dots $, which had  natural geometric
transformation rules.
The first of these gauge fields,
$ g_{(2)}^{\mu \nu }$, is the inverse of the usual world-sheet metric. Just as
in
Riemannian geometry one can construct the tensor density
$ \tilde g_{(2)}^{\mu \nu }$ in terms of the tensor
$ g_{(2)}^{\mu \nu }$ by
$ \tilde g_{(2)}^{\mu \nu }=\sqrt{g} g_{(2)}^{\mu \nu }$, it seems that in
\W-geometry, a \W-density can be constructed from a \W-scalar, and the
\W-scalar
can be thought of as giving  some kind of generalisation
of the Riemannian line
element [\hegeoma]. Further, \W-scalars can be constructed in any dimension,
but
\W-densisties and hence invariant actions can only be constructed in one and
two
dimensions [\hegeoma].

So far, only the case in which the matter system
is a single boson has been
discussed. It is clearly important to
generalise the results given here to less
trivial matter systems. It turns out that it is straightforward although
non-trivial to  generalise to the case of multi-boson realisations. For
$n$ bosons $\phi ^i$, $i=1,\dots ,n$, one obtains a construction
on a bundle with local coordinates $x^\mu, y_\mu^i$ and fibres $(T^*\N)^n$
but the gauge group remains essentially the same
as in the one boson case;
further details will be given elsewhere.

One motivation for the study of \W-geometry is to try to understand finite
\W-transformations (as opposed to
those with infinitesimal parameters) and the
moduli space for \W-gravity. The infinitesimal transformations for the scalar
field $\phi$ were derived from studying infinitesimal symplectic
diffeomorphisms and it follows that the large W-transformations of $\phi$ are
given by the action of large symplectic diffeomorphism
 transformations on $y_\mu =
\partial _\mu \phi$.   It seems natural to conjecture that the
transformations of the gauge fields can be defined to give invariance under
the full group of symplectic diffeomorphisms, as opposed to invariance under
the subgroup   generated by exponentiating infinitesimal ones, but this
remains to be proved.

Another important issue is that of quantum \W-gravity. Here
we shall briefly review the linearised
conformal gauge results of [\wanog]; see [\vanlig] for a discussion
of light-cone gauge \W-gravity, including non-linear corrections.
We have seen that
 for each spin $s$, $2 \le s \le N$, there is a symmetric tensor
gauge field $g^{(s)}_{\mu _1 \mu _2 \dots \mu _s}$ transforming under
linearised \W-gravity  and   \W-Weyl transformations as
$$
\delta g^{(s)}_{\mu _1
\mu _2 \dots \mu _s}= \partial _{(\mu _1}
\lambda ^{(s)}_{\mu _2 \dots \mu _s )} +\eta_{(\mu _1
\mu _2}\sigma_{\mu _3 \dots \mu _s )} +O(h)
\eqn\retert$$
The \W-gravity parameter $\lambda ^{(s)}_{\mu _1 \dots \mu _{s-1}}$ is
a rank-$(s-1)$ symmetric tensor and the \W-Weyl parameter
$\sigma_{\mu _1 \dots \mu _{s-2}}$
is
a rank-$(s-2)$ symmetric tensor.
Note that the part of the transformation
\retert\ involving the
trace of
$\lambda ^{(s)}_{\mu _1 \dots \mu _{s-1}}$ can be absorbed into a
redefinition of the \W-Weyl parameter.
The linearised spin-$s$ curvature
$$\eqalign{
R^{(s)}_{\mu_1 \nu_1 \ \mu_2 \nu_2 \, \dots \, \mu _s \nu _s}
&=\Bigl( \bigl[ \{ \partial _{\mu _ 1}\partial _{\mu _ 2}\dots \partial
_{\mu _ s} g^{(s)}
_{\nu _1 \nu _2 \dots \nu _s}
- (\mu _1 \leftrightarrow \nu _ 1) \}
\cr &
- (\mu _2 \leftrightarrow \nu _2 ) \bigr] \dots
- (\mu _ s\leftrightarrow \nu _ s)\Bigl)
+O(g^2)
\cr}
\eqn\curw$$
is invariant under \retert\ to lowest order in the gauge fields.
It has $s$ anti-symmetric pairs of indices and is symmetric under the
interchange of any two pairs. The corresponding linearised
curvature scalar is given
by
$$
R^{(s)}
={1 \over 2^s} \epsilon ^{\mu_1 \nu_1}\dots \epsilon ^{\mu _s \nu _s}
R^{(s)}_{\mu_1 \nu_1 \,
\dots \, \mu _s \nu _s}
= \epsilon ^{\mu_1 \nu_1}\dots \epsilon ^{\mu _s \nu _s}
\partial _{\mu _ 1}\partial _{\mu _ 2}\dots \partial
_{\mu _ s} g^{(s)}
_{\nu _1 \nu _2 \dots \nu _s}
\eqn\tert$$
The effective
action  obtained by integrating out the matter takes the form
$$
\Gamma =\sum _{s=2} ^N a_s \ix R ^{(s)}{1 \over \square }R^{(s)}
\eqn\rorthy$$
for some constants $a_s$.
This is invariant, at least to lowest order in the gauge fields, under the
\W-gravity transformations
but not under the \W-Weyl transformations:
$$
\delta \Gamma=-2\sum _{s=1}^{N} \ix \bigl[  a_{s} R^{(s)}
\epsilon ^{\mu_1 \nu_1} \dots  \epsilon ^{\mu_{s-2} \nu_{s-2}}
\partial_ {\mu_1} \dots \partial_ {\mu_{s-2}}
{\sigma _{\nu _1   \dots \nu _{s-2}  }}\bigr]
\eqn\erte$$
Thus, at the linearised level,
there are no \W-gravity anomalies but there are \W-Weyl anomalies.

In general, the gauge symmetries can be used to gauge away
 all but the total traces of the gauge fields (modulo some subtleties
 discussed in [\wanog]). For even spins $s=2r$, this
  leaves
a scalar field $\rho^{(2r)} \propto
{g^{(2r)}_{\nu _1 \dots \nu _{ r}}}^{\nu _1 \dots \nu _{ r}}$,
while for
odd spins $s=2r+1$ this leaves
  a vector gauge field $A ^{(2r+1)}_{\mu }\propto
{g^{(2r+1)}_{\mu \nu _1 \dots \nu _{ r}}}^{\nu _1 \dots \nu _{ r}}$
with field strength $F^{(2r+1)}= \ep \dm A ^{(2r+1)}_{\nu }$.
 The residual spin-one gauge invariances can then be fixed using
the Lorentz gauge condition $ \partial ^\mu A ^{(2r+1)}_{\mu
}=0$, which can be solved to give
$A ^{(2r+1)}_{\mu
}=\epsilon _{\mu \nu} \partial ^\nu \rho^{(2r+1)}$ (plus a zero-mode piece)
for some scalar $\rho^{(2r+1)}$.
Thus all of the \W-symmetries  are fixed by the gauge choices
$$\eqalign{
g^{(2r)}_{\mu _1 \dots \mu _{2r}}&=(-1)^{ r}
\eta_{(\mu _1 \mu _2 }\dots \eta_{\mu _{2r-1} \mu _{2r} )}
\rho^{(2r)}
\cr
g^{(2r+1)}_{\mu _1 \dots \mu _{2r} \nu}&=(-1)^{ r}
\eta_{(\mu _1 \mu _2 }\dots \eta_{\mu _{2r-1} \mu _{2r} }
\epsilon _{\nu )\lambda} \partial ^ \lambda \rho^{(2r+1)}
 \cr}
\eqn\gaga$$
and
the linearised curvature scalars \curw,\tert\
become
$$
R^{(2r)}= {\square}^r \rho ^{(2r)}
,\qquad
R^{(2r+1)}= {\square}^r F^{(2r+1)}={\square}^{r+1} \rho ^{(2r+1)}
\eqn\erht$$
 Substituting this in \rorthy\ gives the effective action
$$\Gamma =\sum _{r=1}^{[N/2]} \left[ a_{2r} \rho
^{({2r})}{\square}^{{2r}-1}\rho
^{({2r})} +a_{2r+1}F^{(2r+1)}\square ^{2r-1} F^{(2r+1)} \right]
=
\sum _{s=2}^N a_s \rho ^{(s)}{\square}^{s-1}\rho ^{(s)}
\eqn\actre$$
consisting of the Liouville action for $s=2$ and higher derivative
counterparts for the higher spin cases. It is perhaps surprising that
the \W-gravity generalisation of
Liouville theory that emerges in this approach
is not the Toda theory advocated in [\manspen] and elsewhere, but a higher
derivative version of this. (Note, however, that an alternative, but
less natural, geometric framework
that does lead to   Toda-theory version of \W-gravity was given in [\wanog];
it is related to the one described
here by a non-local change of variables.)

 Consider now
the moduli space $M_n$ for gauge fields $\tilde g_{(n)}$ subject to the
constraints generated by \decon\ (and the $\W_N$ constraint \noo, if
appropriate) [\wanog]. Linearising about a Euclidean background $\tilde F =
\half
\tilde g_{(2)} ^{\mu \nu}y_{\mu }y_{\nu}$ and choosing complex coordinates
$z,\bar z$
on the Riemann surface $\N$ such that
the background is $\tilde F= y_z y_{\bar z}$,
 and using  the linearised transformations $\delta  \tilde g_{(n)}^{zz \dots
z}= \partial _{\bar z} \lambda _{(n)}^{zz \dots z}$,
it follows by standard
arguments that the tangent space to the moduli space
$M_n$ at a  point
corresponding to the background configuration is the
space of holomorphic
$n$-differentials, \ie\ the $n$-th rank symmetric
tensors $\mu _{zz \dots z}$
with $n$ lower $z$ indices satisfying
$\partial _{\bar z} \mu _{zz \dots z}=0$
[\wanog]. It follows from the Riemann-Roch
theorem that the dimension of this
space on a genus-$g$ Riemann surface (the
number of anti-ghost zero-modes) is
$dim(M_n)= (2n-1)(g-1)+k(n,g)$ where $k(n,g)$
is the number of solutions
$\kappa ^{zz \dots z}$ (with $n-1$  \lq $z$' indices) to $\partial _{\bar
z}\kappa ^{zz \dots z}=0$ (the number of ghost
zero-modes). It would  be of
great interest to use information about the
global structure of the symplectic
diffeomorphism group to learn more about the
structure of these moduli spaces.

 \refout
\end